\documentclass[11pt]{article}

\usepackage[mathscr]{eucal}
\usepackage{%txfonts,pxfonts,
amsmath,amsfonts,amssymb,amsthm,bm,calrsfs,stmaryrd}
\usepackage{hyperref}
\usepackage{times}
\usepackage{graphicx}
\usepackage{listings}
\usepackage[title,titletoc,toc]{appendix}

\usepackage[T1]{fontenc} 
\usepackage[latin1]{inputenc}
\usepackage[english]{babel}
\usepackage{amsmath,amsthm,amssymb}
\usepackage{pifont}
\usepackage{tikz}
\usepackage{dsfont}\let\mathbb\mathds
 \usepackage{footnote} 
\usepackage{pslatex}

\usepackage{amsmath,amsbsy,amsfonts,amssymb,amsthm,dsfont,fullpage,units}
\usepackage{graphicx}
\usepackage{authblk}
\usepackage{algorithm,algorithmic}
\usepackage{color}
\usepackage{tikz}
\usetikzlibrary{calc,shapes}

\usepackage{caption}
\usepackage{subcaption}

\title{Spectral Sequence Motif Discovery}
\author[1]{Nicol\'o Colombo}
\author[2]{Nikos Vlassis}
\affil[1]{nicolo.colombo@uni.lu, Luxembourg Center for Systems Biomedicine, University of Luxembourg}
\affil[2]{vlassis@adobe.com,  Adobe Research, San Jose, CA\thanks{The bulk of this work was carried out when NV was with the Luxembourg Centre for Systems Biomedicine.} }

\begin{document}
\maketitle

\begin{minipage}{.90\textwidth}

\textsc{Abstract.}
Sequence discovery tools play a central role in several 
fields of computational biology.
In the framework of Transcription Factor binding studies, 
motif finding algorithms of increasingly high performances 
are required to process the big datasets 
produced by new high-throughput sequencing technologies.
Most existing algorithms are computationally demanding 
and often cannot support 
the large size of new experimental data.
We present a new motif discovery algorithm that is built on a recent 
machine learning technique, referred to as Method of Moments.
Based on spectral decompositions, this method  
is robust under model misspecification and 
not prone to locally optimal solutions.
We obtain an algorithm that is extremely fast and designed 
for the analysis of big sequencing data.
In few minutes, 
we can analyse datasets of more than hundred thousand sequences 
and produce motif profiles that match those computed 
by various state-of-the-art algorithms.

\end{minipage}

\vspace{15pt}

\section{Introduction}

In the last decades, due to the advent  of new sequencing technologies, 
motif discovery algorithms have become an essential tool in many computational 
biology fields.
In cell biology, sequence motif discovery plays a primary role in the 
understanding of gene expression
through the analysis of sequencing data and the identification of 
DNA-transcription factors binding sites
\cite{berger_compact_2006,wei_genome-wide_2010,zhang_simultaneously_2013,
zhao_improved_2012,annala_linear_2011,cheng_computational_2013}.

Various experimental techniques are nowadays available
to extract DNA-protein binding sites in-vivo 
(CHip-Seq \cite{johnson_genome-wide_2007}) and in-vitro  
(PBM \cite{berger_universal_2009}, HT-SELEX 
\cite{tuerk_systematic_1990,kinzler_gli_1990,jolma_multiplexed_2010,
jolma_dna-binding_2013}).
Thanks to the quantity and quality of data produced, 
HT-SELEX is considered one of the most promising high-throughput  
techniques for studying transcription factors binding affinity
in-vitro (see the recent work \cite{orenstein_comparative_2014} 
for a quantitative comparison between HT-SELEX 
and other high-throughput techniques as CHip-seq and PBM).  
In the HT-SELEX protocol, tens of thousands enriched 
DNA fragments are obtained through a series of incubation/selection cycles.
In each cycle, an initial pool containing randomised ligands of length 14-40 bp
is incubated with an immobilised DNA-binding protein.
Bounded ligands are amplified by PCR, sequenced and 
then used as initial pool for a next cycle, until the pool
is saturated \cite{tuerk_systematic_1990,kinzler_gli_1990,zhao_inferring_2009,
jolma_multiplexed_2010,jolma_dna-binding_2013}.

Due to the high but not exact specificity of transcription factors 
binding affinities, 
enriched DNA fragments in a dataset typically contain similar but not
exactly conserved instances of the same binding motif.
Sequence discovery algorithms are required to produce 
binding models  that are at the same time intuitively clear and 
able to capture the full complexity of such probabilistic 
mechanisms \cite{zhao_inferring_2009,
jolma_multiplexed_2010,berger_universal_2009}.  
In the simplest case, binding preferences are reported using consensus 
sequences, obtained by selecting a few deterministic character strings 
that are over-represented in the dataset.
A more flexible representation is provided by Position Weight Matrices (PWM) 
that describe binding sites as probability densities 
over the DNA alphabet.
Based on the stringent assumption that 
the total binding energy is a site-by-site sum of single 
protein-nucleobase interactions,
PWM's are only approximate models of the 
true transcription factor preferences.
A debate is still open on whether such  
approximation gives a satisfactory picture  
of the DNA-proteins interaction
or is a too simplified reduction of the real biological 
process \cite{badis_diversity_2009}.
More sophisticated models, that go beyond the PWM representation
by taking into account multiple bases probability distributions
or long-distance interactions, 
have been proposed and tested in the literature 
\cite{bulyk_nucleotides_2002,chen_rankmotif++:_2007,
badis_diversity_2009,santolini_beyond_2013,mathelier_next_2013}.
However, in most cases, these improvements could not provide 
any evidence against the simpler and more intuitive approach 
based on position independent distributions \cite{zhao_quantitative_2011}.

In machine learning, factorized (aka product) distributions 
like PWMs and their linear combinations (aka mixtures) are 
commonly used in modelling empirical distributions 
from various kinds of data.
The problem of learning mixtures of product distributions from 
given datasets has been intensively studied \cite{titterington_statistical_1985,
lindsay_mixture_1995}.
In particular, as proposed by Chang (1996),
it is possible to infer a mixture of product distributions
via the spectral decomposition of `observable' matrices, \emph{i.e.}
matrices that can be estimated directly from the data using suitable
combinations of the empirical joint probability distributions \cite{chang_full_1996}.
Extensions and improvements of this idea have been 
developed more recently in a series 
of remarkable works, where the spectral technique is applied to a 
larger class of probability distributions, 
and robust versions of the original method have been analysed
theoretically \cite{mossel_learning_2006,anandkumar_method_2012,
hsu_spectral_2012,anandkumar_tensor_2012}.

In this paper, we further develop the original 
spectral technique \cite{chang_full_1996} and study its 
application to the 
problem of learning probabilistic motif profiles from noisy sequencing data.
The key observation is that, in the PWM approximation, 
motif discovery reduces to the more general issue 
of learning a mixture of product distributions 
and hence it is possible to extract motif profiles from 
sequencing data using usual spectral decompositions.
In particular, we combine some of the improvements introduced in 
\cite{mossel_learning_2006} and \cite{corless_reordered_1997,anandkumar_tensor_2012} 
to obtain a more stable spectral decomposition and implement 
a few new ideas to adapt the general technique to the DNA sequence discovery problem.  
Through this work, we always assume that transcription factors specificities 
are well described by product distributions \emph{i.e.} PWM's,     
and leave for future work the spectral   
inference of more advanced motif representations.   

The paper is organised as follows: in Section \ref{related sec} we give a brief overview of some related works, in Section \ref{method sec} we 
describe the spectral methods and their applications to sequencing datasets and in Section \ref{result sec} we summarize our main results.
More mathematical details about the spectral techniques can be found in in 
Appendix \ref{spectral appendix} and a schematic 
version of our algorithm is provided in Appendix \ref{algorithm appendix}.
A software implementation of our method is available under request at \texttt{nicolo.colombo@uni.lu}.

\section{Related Work}
\label{related sec}

\paragraph{Motif Finding} 
The literature on sequence motif discovery is vast.
We refer to \cite{tompa_assessing_2005,das_survey_2007,
sandve_improved_2007,simcha_limits_2012} for reviews and additional references.
There are two main classes of motif finding algorithms,
probabilistic and word-based.
Probabilistic algorithms search for the most represented 
un-gapped alignments in the sample to obtain 
deterministic consensus sequences, PWM models, or more advanced models 
that take into account multi-base correlations \cite{bulyk_nucleotides_2002,chen_rankmotif++:_2007,
badis_diversity_2009,santolini_beyond_2013,mathelier_next_2013}.
Word-based algorithms search the dataset
for deterministic short words, measure the statistical significance 
of small variations 
from a given seed, or transform motif discovery into a kernel feature classification 
problem \cite{leslie_spectrum_2002,vert_kernels_2005,lee_discriminative_2011}. 
Our method and two of the algorithms we have used for evaluating our 
results, namely MEME \cite{bailey_fitting_1994} and STEME 
\cite{reid_steme:_2011}, belong to the probabilistic class, while the method used in 
\cite{jolma_dna-binding_2013} and DREME 
\cite{bailey_dreme:_2011} are word-based algorithms.
The latter algorithms can also compute PWM models, so it is of interest to compare 
algorithms of different classes (See Results section).    

\paragraph{Spectral Methods}
Spectral methods have been applied as an alternative to the Expectation 
Maximization algorithm \cite{dempster_maximum_1977} for inferring 
various kinds of probability distributions, such as mixtures of product 
distributions, Gaussian mixtures, Hidden Markov models, and others \cite{chang_full_1996,mossel_learning_2006,
hsu_spectral_2012,anandkumar_tensor_2012,anandkumar_learning_2012,anandkumar_tensor_2012,boots_closing_2011} 
(see \cite{balle_methods_2014} for a recent review).
These methods are not as flexible 
as the Expectation Maximization algorithm, but they are not prone to local 
optima and have polynomial computational time and sample complexity.
Various spectral decomposition techniques  have been proposed:  
Chang's spectral technique of 
\cite{chang_full_1996,mossel_learning_2006}, 
the symmetric tensor decomposition 
presented in \cite{anandkumar_tensor_2012}, 
and an indirect learning methods for inferring the parameter of Hidden 
Markov Models \cite{hsu_spectral_2012}.   
The practical implementation of the spectral idea is a nontrivial task  
because the stability of spectral decomposition strongly depends on the spacing between the eigenvalues of the empirical matrices. 
In \cite{mossel_learning_2006,anandkumar_tensor_2012} certain eigenvalue
separation guarantees for  Chang's 
spectral technique  are obtained via the contraction 
of the higher (order three) moments to Gaussian random vectors.
In the tensor approach presented in \cite{anandkumar_tensor_2012}, the 
non-negativity of the eigenvectors is 
ensured by using a deflating power method that generalizes 
usual deflation techniques for matrix diagonalization to the case of 
symmetric tensors of order three. 
A third possibility involves replacing the random vector of  
Chang's spectral technique with an  `anchor observation' that, 
for each hidden state, `tends to appear in the state much more often 
than in the other states' \cite{song_spectacle:_2014} and guarantees the presence of at least one  well separated 
eigenvalue \cite{arora_learning_2012,song_spectacle:_2014}. 
Finally, as briefly mentioned in 
\cite{anandkumar_tensor_2012,hsu_learning_2012}, the stability of Chang's technique can be significantly improved through the simultaneous 
diagonalization of several random matrices.
Here, we present a new approach based on the simultaneous 
Schur triangularization of a set of nearly commuting 
matrices \cite{corless_reordered_1997}.

\paragraph{Spectral Method and Sequence Analysis}
To the best of our knowledge, spectral 
methods have not been applied so far to the problem of DNA sequence motif discovery that 
we address here.
Nevertheless, spectral techniques have been applied to other 
types of sequence analysis problems, 
such as poly(A) motif prediction \cite{xie_poly_2013},
chromatin annotation \cite{song_spectacle:_2014}, 
and sequence prediction \cite{quattoni_spectral_2014}. 
The techniques used in these works are all based, with minor 
modifications, on the spectral algorithm of Hsu et al.~\cite{hsu_spectral_2012} 
for learning Hidden Markov Models,
in which a dataset of time-series of observed values 
$\{x_1,x_2,\dots \}$ is used to recover a single 
observation matrix ($O_x$)  
whose columns are the conditional probabilities  
associated with the hidden states.
Our approach 
marks a significant departure from these methods by allowing the 
recovery of distinct observation matrices
($O_x,O_y,\dots$) and hence the extraction of motif PWM's.
Finally, we note that a general technique for learning mixture of 
product distributions in the presence of a background has been recently 
presented \cite{zou_contrastive_2013}; it would be interesting 
to study how this technique could be applied to the problem of 
sequence motif discovery.

\section{Methods}
\label{method sec}

DNA-protein interactions can be approximated by PWM models 
under the assumption that the total binding energy is the sum of single protein-nucleobase interactions.
In this case, Transcription Factors binding affinities are represented by
$d\times \ell$ frequency matrices, where $d$ is the dimensionality
of the sequences alphabet and $\ell$ the length of the binding site.
Looking at these frequency matrices as components
of a mixture of product distributions, we  recover all their entries 
via  the spectral decomposition of  "observable" matrices
computed from data.

To obtain the empirical distributions 
we use a length-$\ell$ sliding window
that runs over all sequences in the dataset with one-character steps.
We take into account the possible presence of secondary motifs
by considering mixtures with high number of components.
This choice is also motivated by the fact that 
the one-character steps of the sliding window can produce strong signals 
for many shifted versions of the same sub-sequences. 
In practice, if $p$ is the number of components in the mixture 
we set $p>15$ and select the $p_{\rm top}\sim 3$ most informative components at the 
end, according to their relative entropy respect to a background distribution.
Moreover, since spectral techniques do not apply when the number of mixture components 
is higher than the dimension of the sequence alphabet,
we augment the size of the sequences space by 
grouping contiguous variables and work with an alphabet of higher dimensionality.
Concretely, letting $d$ be the dimensionality of the sequence alphabet ${\cal A}$ 
and $n$ the number of variables in a group, the corresponding  
grouped variables have dimensionality $d^n$  and take values in the alphabet ${\cal A}^{\otimes n }$.
For example, if $x_1, \dots x_\ell$ are the single character variables 
of a length-$\ell$ sliding window $W=[x_1, \dots, x_{\ell}]$, we consider three grouped variables 
 $x=[x_1, \dots, x_n] $, $y=[x_{n+1},\dots ,x_{2n}]$, $z=[x_{2n+1},\dots, x_{\ell}]$
 such that $W=[x,y,z]$.   

Assuming for simplicity a mixture of product distributions 
with $p=d^n$  hidden  components 
defined over a $d^n$-dimensional space,
pairwise and triple probability tensors read  
\begin{equation}
\label{joint probability}
[P(x,y)]_{ij} =   \sum_{r=1}^{p} h_r \ X_{ir} Y_{jr} , \qquad   
[P(x,y,z)]_{ijk} = \sum_{r=1}^{p} h_r  \ X_{ir} Y_{jr} Z_{kr}  \qquad i=1,\dots,d^n
\end{equation}
where  
$[P(x)]_i$ is the probability of observing $x=i$,
the mixing weights $h_r$ satisfy $0<h_r<1$  and $\sum_r h_r=1$ 
and $X, Y, Z$ are $d^n\times p$ matrices 
that contain 
the conditional probability distributions, for variables $x,y,z$ respectively.
For any $d^n$-dimensional vector $\theta$ one has
\begin{equation}
\label{contracted probability}
[P_{\theta}(x,y,z)]_{ij} =\sum_k [P(x,y,z)]_{ijk} \ \theta_k ~=~ [X {\rm diag}(h) {\rm diag}(\theta^T Z ) Y^T]_{ij}
\end{equation}
where ${\rm diag}(h)$ and ${\rm diag}(\theta^T Z )$ are $p\times p$ diagonal matrices whose entries are the components 
of  $h=[h_1, \dots h_{p}]$ and  $\theta^T Z$ respectively.
If the conditional
probability matrices $X,Y$ have both rank $p$  and $h_r>0$  for all $r=1,\dots,p$, 
from \eqref{joint probability} and  \eqref{contracted probability} one obtains the matrix 
\begin{equation}
\label{chang matrix}
S(\theta)~=~ P_{\theta}(x,y,z) \left( P(x,y) \right)^{-1} ~=~ X {\rm diag}(\theta^T Z ) X^{-1}  
\end{equation}
that is called `observable'  because its empirical estimation, say $\hat S(\theta)$, can be directly obtained from 
the sequences sample, using the joint empirical probabilities $\hat P(x,y)$ and $\hat P(x,y,z)$. 
When the dimensionality of the sequence alphabet $d$ and the 
number of grouped variables $n$ is big, the manipulation of the
$d^n \times d^n$ empirical matrices can be computationally expensive.
Moreover, it can be useful in general to learn a mixture with a smaller 
the number of mixture components $p<d^n$.
This is obtained by reducing  
$\hat S(\theta)$ down to a $p\times p$ matrix through  
a rank-$p$ approximation of the empirical distributions.
More precisely, all empirical 
moments become $p\times p$ matrices after multiplication (from the left and from the right) with  the transpose of 
suitable $d^n\times p$ rectangular matrices, formed with   
the first $p$ (left and right) singular vectors   
of the empirical pairwise probabilities $\hat P(x,y)$ and $\hat P(x,z)$.
See Appendix \ref{spectral appendix} for a formal definition of 
$\hat P(x,y)$  and $\hat P(x,y,z)$ and more details on their rank-$p$
reduction.

In the case of misspecified models, \emph{i.e.} when the sample 
is not drawn 
exactly from a mixture of product distributions with exactly $p$ 
mixture components, $\hat S(\theta)$ can  be 
identified only approximately with the right hand side of \eqref{chang matrix}.
Even in this case,  it can be shown that the recovery of $X$ and $Z$ 
is theoretically possible, with certain success guarantees \cite{mossel_learning_2006,
hsu_spectral_2012,anandkumar_method_2012}.
However, when the empirical moments are not exact, the stability of the spectral method
depends strongly on the (real) separation between the eigenvalues
of  $\hat S(\theta)$.
A theorem, \emph{ Lemma 4 (Eigenvalues separation)} in \cite{mossel_learning_2006},  proves  
that a sufficient separation is obtained with
high probability if $\theta$ is chosen to be a random vector whose entries are 
independent Gaussians with mean 0 and variance 1.
To increase further the stability  of our algorithm,   
we define a set of distinct random vectors $\{\theta_1, \theta_2, \dots\}$
and simultaneously diagonalise 
the set of (nearly commuting)  empirical matrices $\hat S(\theta_i)$.
As suggested in \cite{corless_reordered_1997}, if $\hat S(\theta_i)$ for $i=1,2 \dots$ is a set of nearly commuting
matrices and $Q,\sigma$ are respectively the orthogonal and upper triangular matrices 
in the Schur decomposition of their linear combination 
$\hat S=\sum_i \hat S(\theta_i) =Q\sigma Q^T$, the approximate eigenvalues of 
each $\hat S(\theta_i)$ can be read from the diagonal of $\tilde T_i=Q^T \hat S(\theta_i) Q$.
Since the matrices $\hat S(\theta_i)$ do not commute exactly, the triangularisation obtained via $Q$ is
approximate and $\tilde T_i$ 
can contain small entries below the diagonal.
It can be shown (see Appendix for details) that the size of such entries is proportional to $\varepsilon=\max_{i,j} \| \hat S(\theta_i)\hat S(\theta_j)- \hat S(\theta_j)\hat S(\theta_i) \|_F$, where $\|A\|_F=\sqrt{{\rm Tr}[A^T A]}$, provided that the eigenvalues of $\hat S$ are well separated.
The error in the eigenvalues estimation can then be bounded via usual eigenvalues 
perturbation theorems, where the norm of the perturbation matrix is proportional to $\varepsilon$.   

Thus, we choose a set of $d^n$-dimensional random vectors $\{\theta_1, \dots, \theta_p \}$ and 
form a $p\times p$ matrix $\Lambda$ whose rows contain the approximate eigenvalues of the corresponding matrices $\hat S(\theta_i)$.
The matrices Z is then computed from $Z=\Theta^{-1} \Lambda$, where $\Theta$ is a $p\times d^n$ matrix formed with the  
$d^n$-dimensional random vectors $\{\theta_1, \dots, \theta_p \}$.
Finally we recover  $X$ and $Y$ using the obtained $Z$ to approximately diagonalise sets of analogous "observable" 
matrices, say  $\hat S_x(\theta_i) $ and  $\hat S_y(\theta_i) $, 
obtained from different combinations of the empirical moments, see Appendix \ref{spectral appendix} for more details. 

Together, the 
matrices $X,Y,Z$ combine to build a set of higher dimensional PWM's defined by
\begin{equation}
H_r \in [0,1]^{d^n\times 3} \qquad s.t. \qquad \left\{ \begin{array}{l} [H_r]_{i1}=X_{ir} , \qquad [H_r]_{i2}=Y_{ir} , \qquad [H_r]_{i3}=Z_{ir}\\ \sum_i[H_r]_{ik} =1 \qquad \forall k=1,2,3\end{array} \right.
\end{equation}
for $r=1,\dots, p$ and $i=1,\dots d^n$. 
To select the nontrivial mixture components $H^{\rm top}_r$ we compute, for each $H_r$, the relative entropy 
$I(r)=\sum_{j,k}[\frac{H_r}{B}]_{ij} \log [\frac{H_r}{B}]_{ik}$, where $B$  is a background distribution
obtained form a control dataset.
In particular, the control dataset consists of sequences that come from a different transcription factor experiment
and the background distribution is obtained  from its empirical joint probability 
distributions (see Appendix \ref{spectral appendix} for more details).
We choose the models of smallest relative entropy 
to create sub-sequences  alignments and extract the final $d$-dimensional models $h^{\rm top}_r$.
For each $r=1,\dots,p_{\rm top}$, we define a scoring function 
using the log-likelihood of $H^{\rm top}_r$ and include in the alignment 
all length-$\ell$ sub-sequences whose score is above a certain threshold.
The thresholds are chosen within a finite set of possible values\footnote{We 
have used this heuristic approach since, 
in the case of probabilistic models, the problem
of choosing an optimal matching threshold given a set of sequences 
has been proven to be NP-hard \cite{touzet_efficient_2007}.}, in order  to maximise 
the small-sample corrected information content of the low-dimensional model 
$h^{\rm top}_r$.

\section{Results}
\label{result sec}

\begin{figure}[t]
\centering
\begin{tabular}{llccr}
Jolma et al. \cite{jolma_dna-binding_2013} &
\includegraphics[scale=0.42,trim=0 3mm 0 0,clip=true]{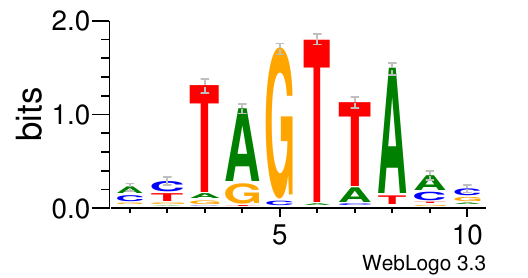}&
\includegraphics[scale=0.42,trim=0 3mm 0 0,clip=true]{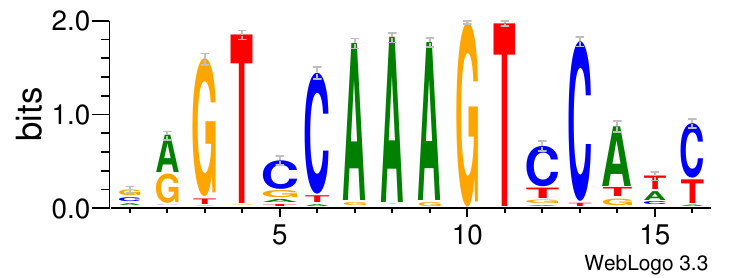}&
\includegraphics[scale=0.42,trim=0 3mm 0 0,clip=true]{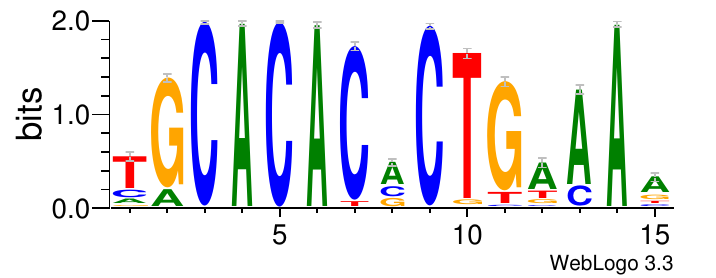}&
\includegraphics[scale=0.42,trim=0 3mm 0 0,clip=true]{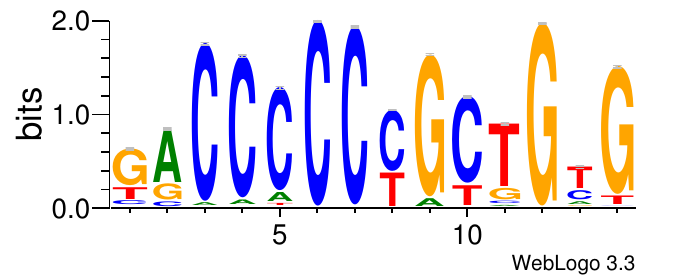}\\
\\
Our Method &
\includegraphics[scale=0.42,trim=0 3mm 0 0,clip=true]{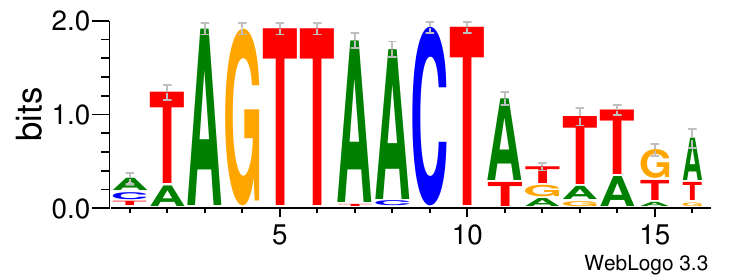}&
\includegraphics[scale=0.42,trim=0 3mm 0 0,clip=true]{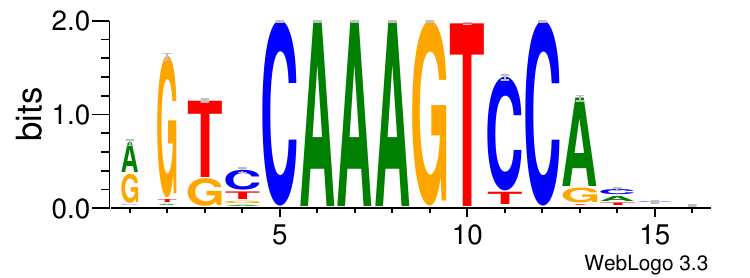}&
\includegraphics[scale=0.42,trim=0 3mm 0 0,clip=true]{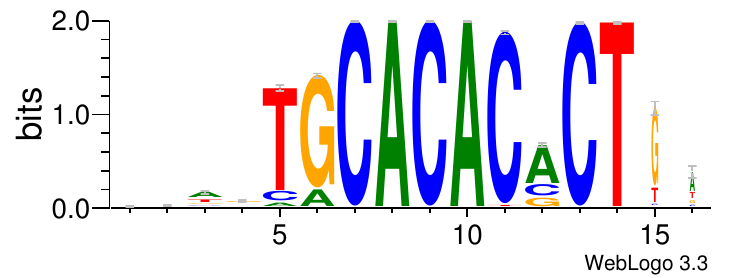}&
\includegraphics[scale=0.42,trim=0 3mm 0 0,clip=true]{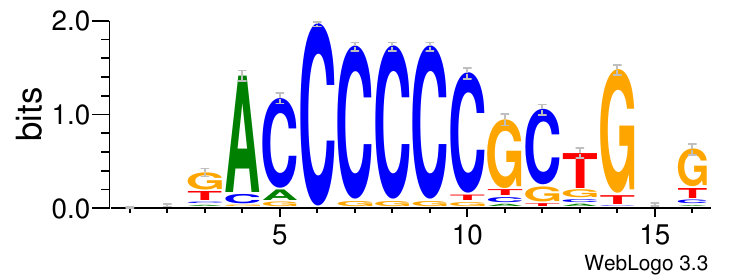}\\
\end{tabular}
\caption{HMBOX1, HNF4A, ZSCAN4 and ZIC1 models 
published in \cite{jolma_dna-binding_2013}  and models 
computed using our method on the same datasets.}
\label{more logo pic}
\end{figure}

We present a new motif finding algorithm 
that is faster than other sequence discovery tools
and designed for processing noisy high-throughput
dataset containing more than hundred thousand sequences.
Based on a new and more stable implementation of spectral 
methods \cite{chang_full_1996,mossel_learning_2006,
hsu_spectral_2012,anandkumar_method_2012}, 
our algorithm is robust under model misspecification
and and is not prone to local optima.
Moreover, our algorithm does not require any 
deterministic consensus sequence 
to initialize the search, and models are computed directly
from the empirical joint frequency matrices.
In addition, 
the method is completely general and, upon minor modifications, can be used for
sequence discovery over any sequences alphabet and variable 
number and length of searched motifs, or adapted to analyse 
datasets with binding affinity 
scores \cite{berger_universal_2009,johnson_genome-wide_2007}.

For testing our algorithm we have focused on the transcription factors binding affinity database associated to the recent work: 
\emph{"DNA-Binding Specificities of Human Transcription Facotors"} by Jolma et 
al. \cite{jolma_dna-binding_2013} and
available at the ENA database, under accession number ERP001824.
All datasets consist of $\sim 10^5$  enriched genome fragments of length $\sim 20 $ bp.
For each transcription factor, 
we have download the dataset corresponding to the SELEX cycle 
used in \cite{jolma_dna-binding_2013} to compute the final model
and run our algorithm on it.
Since the amount of ligands with 
specific affinity is expected to increase in each cycle and 
saturate the pool after 4-5 cycle, PWM's are typically extracted from
the third or fourth cycles.

We have used the frequency matrices published in the supplementary
material of \cite{jolma_dna-binding_2013} to evaluate the 
quality of the models computed on the same data using our method.
In Figure \ref{more logo pic} we compare few logos obtained with our algorithm
on the following datasets: HMBOX1 (cycle 4, 29156 sequences ), HNF4A (cycle 4, 80491 sequences), ZSCAN4 (cycle 3, 68378 sequences) and ZIC1 (cycle 3, 267963 sequences).
All logos have been computed using the application \texttt{weblogo 3.3}  
with no options \cite{schneider_sequence_1990}, bash command
\begin{equation}
\label{weblogo}
\texttt{>weblogo -c classic <PWM.txt> PWM.eps} 
\end{equation}
where the file \texttt{PWM.txt} contains the $4\times \ell$ frequency matrix, 
with $\ell$ being the length of the motif.
The sum of the entries in each column equals the number of instances used for 
computing the frequency matrix.
We have chosen this format because the PWM's computed by our algorithm 
and the PWM's published 
in \cite{jolma_dna-binding_2013} are already in this form.

\begin{figure}[t]
\centering
\begin{tabular}{llll}
\vspace{10pt}
Jolma et al. \cite{jolma_dna-binding_2013}&
\includegraphics[scale=0.45,trim=0 3mm 0 0,clip=true]{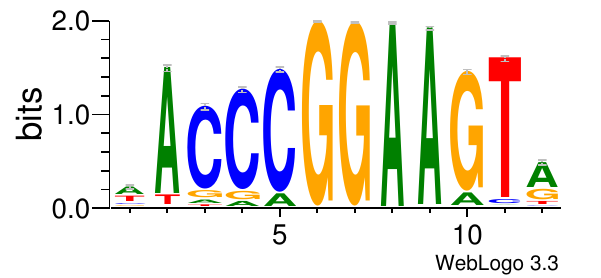}&
\includegraphics[scale=0.45,trim=0 3mm 0 0,clip=true]{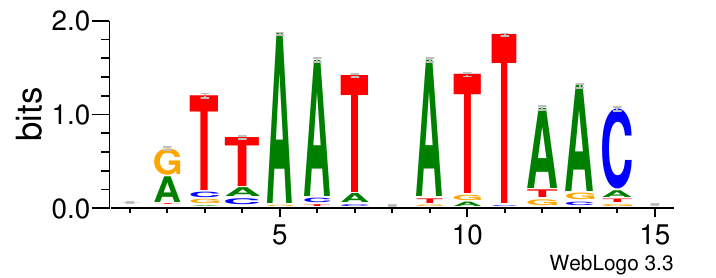}&
\includegraphics[scale=0.45,trim=0 3mm 0 0,clip=true]{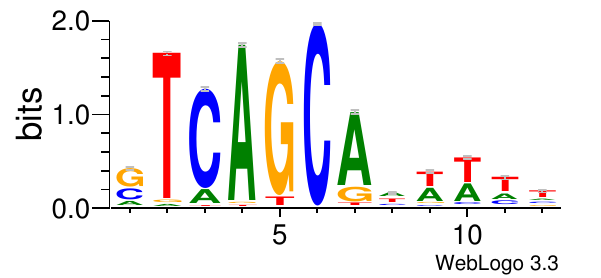}\\
\vspace{10pt}
STEME &
\includegraphics[scale=0.45,trim=0 3mm 0 0,clip=true]{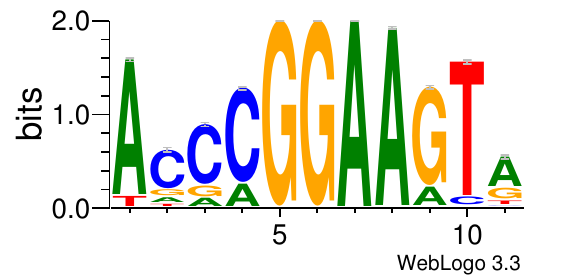}&
\includegraphics[scale=0.45,trim=0 3mm 0 0,clip=true]{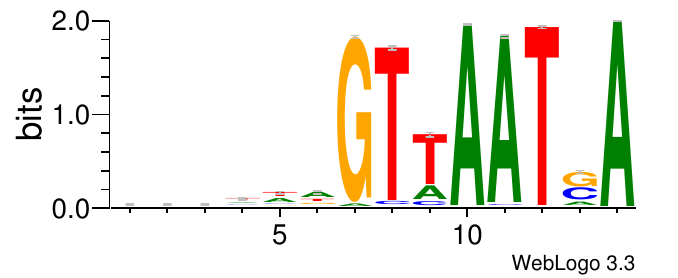}&
\includegraphics[scale=0.45,trim=0 3mm 0 0,clip=true]{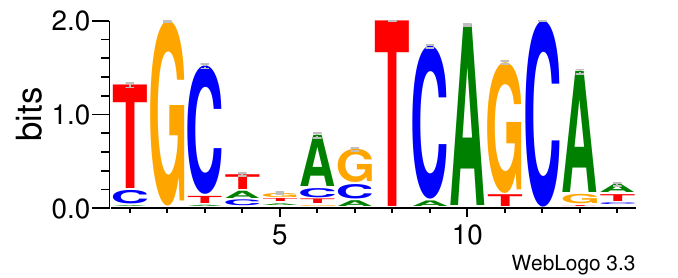}\\
\vspace{10pt}
MEME &
\includegraphics[scale=0.45,trim=0 3mm 0 0,clip=true]{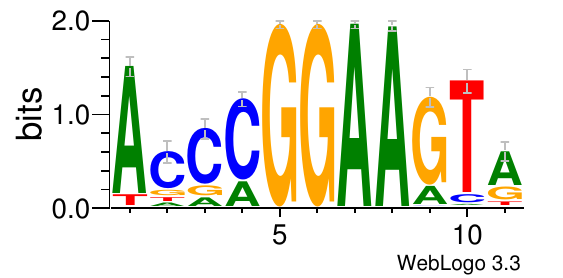}&
\includegraphics[scale=0.45,trim=0 3mm 0 0,clip=true]{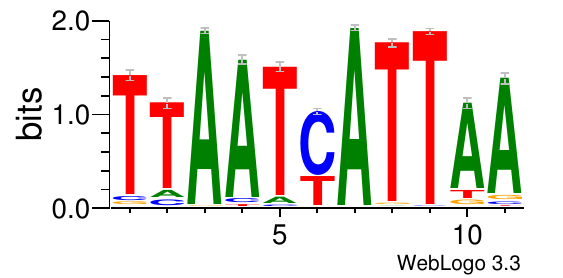}&
\includegraphics[scale=0.45,trim=0 3mm 0 0,clip=true]{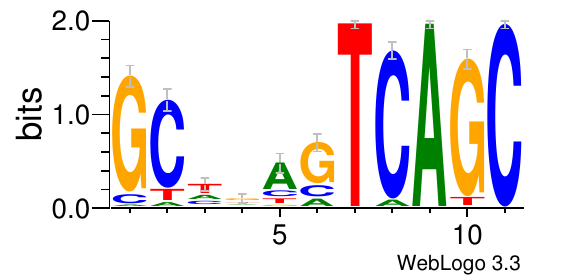}\\
\vspace{10pt}
DREME &
\includegraphics[scale=0.45,trim=0 3mm 0 0,clip=true]{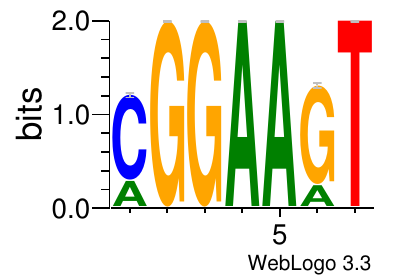}&
\includegraphics[scale=0.45,trim=0 3mm 0 0,clip=true]{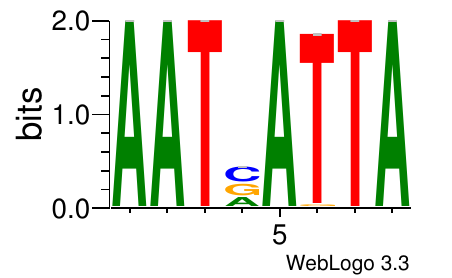}&
\includegraphics[scale=0.45,trim=0 3mm 0 0,clip=true]{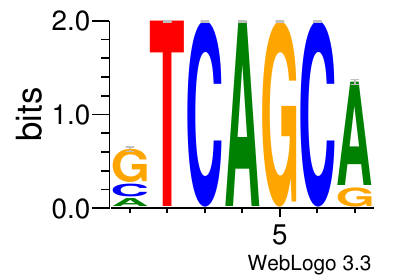}\\
\vspace{10pt}
Our Method &
\includegraphics[scale=0.45,trim=0 3mm 0 0,clip=true]{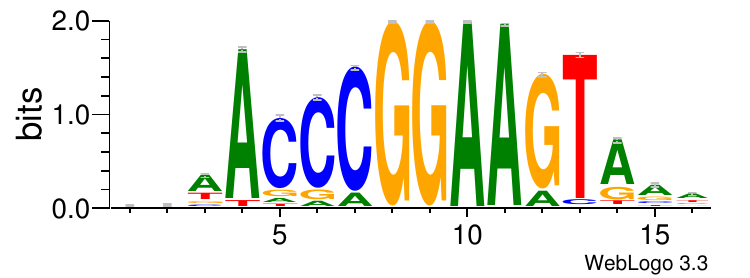}&
\includegraphics[scale=0.45,trim=0 3mm 0 0,clip=true]{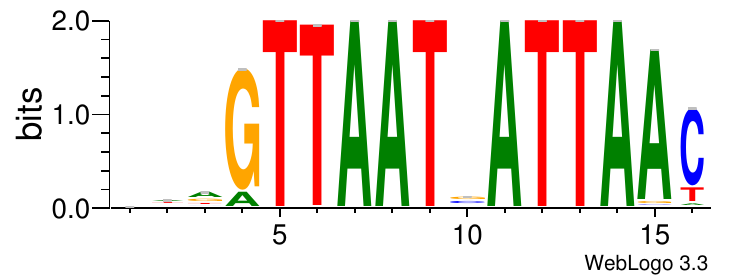}&
\includegraphics[scale=0.45,trim=0 3mm 0 0,clip=true]{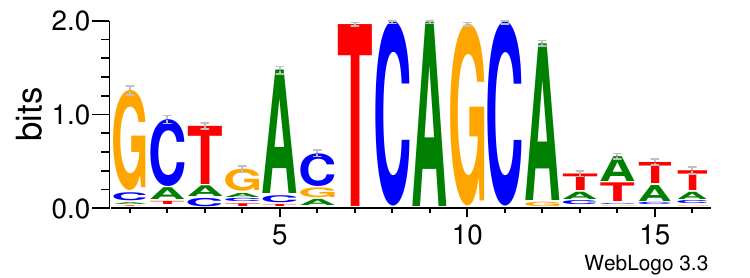}\\
\end{tabular}
\caption{PWM computed by different algorithms on the ELF3, HNF1A and MAFK 
dataset from \cite{jolma_dna-binding_2013}. All logos were obtained using 
\texttt{weblogo 3.3} \cite{schneider_sequence_1990}.}
\label{logos}
\end{figure}

We have then selected three other datasets, ELF3 (cycle 3, 78124 sequences), HNF1A (cycle 3, 142354 sequences) and  MAFK (cycle 3, 144041 sequences),   
and run on each of them few other online available 
algorithms (MEME \cite{bailey_fitting_1994}, DREME \cite{bailey_dreme:_2011},
STEME \cite{reid_steme:_2011}).
All algorithms ran with default settings but 
in many cases we had to reduce the input sequences file 
up to some maximum supported size, 
since none of them could handle files 
of the size of the original datasets.
In figure \ref{logos} we compare the PWM's of 
\cite{jolma_dna-binding_2013} 
with the models obtained by our method (on the whole dataset) and 
by the other algorithms (on a random selection of the maximum supported size).
It should be noted that logos computed on reduced samples 
are in surprisingly
good agreement with the models of \cite{jolma_dna-binding_2013}, computed on the 
whole dataset.   
Since all algorithms we have tried produce more than one motif
for each dataset, we have selected for the comparison the most statistically 
relevant, according to the information provided by the various 
tools on the output page.
Moreover, the PWM's obtained from the online output page of STEME, MEME and  DREME
are Position Specific Probability Matrix (PSPM), whose columns sum to one.
For computing the logos using \eqref{weblogo}, we have transformed the PSPM to frequency matrices  
by multiplying  each entry by the number of sites that were included in the final alignment, 
as it appears on the first line of the output PSPM-format file.
Except for DREME, that produces relatively short motifs ($\ell \le 8$), the 
size of the motifs produced by all algorithms is similar and the typical
range is $10\le \ell \le 14$.

\begin{figure}[t]
\begin{center}
\begin{small}
\begin{tabular}{lcccccc}
     &\vline & DREME & Jolma et al. \cite{jolma_dna-binding_2013} & MEME & STEME & Our Method \\ 
\hline
ELF3 &\vline & 
  0.6171 {\footnotesize ($\pm$ 0.0102)} &   0.6331 {\footnotesize ($\pm$ 0.0116)} &   0.6353  {\footnotesize ($\pm$ 0.0125)} &  0.6367 {\footnotesize ($\pm$  0.0136)} &   0.6373 {\footnotesize ($\pm$ 0.0141)} \\
HNF1A &\vline & 
 0.9818 {\footnotesize ($\pm$ 0.0016)} &    0.9819 {\footnotesize ($\pm$  0.0017 )} & 0.9832  {\footnotesize ($\pm$  0.0009)} & 0.9788  {\footnotesize ($\pm$ 0.0014)} & 0.9794{\footnotesize ($\pm$  0.0019)} \\
MAFK &\vline & 
 0.6007 {\footnotesize ($\pm$  0.0057)}  & 0.5880 {\footnotesize ($\pm$  0.0054)} &  0.6027 {\footnotesize ($\pm$  0.0032)}  &  0.5842 {\footnotesize ($\pm$  0.0036)} &  0.6069{\footnotesize ($\pm$   0.0061)} \\
\end{tabular}
\end{small}
\caption{Average AUC values for the logos shown in Figure 
\ref{logos}. The average and standard variations are computed
over five distinct tests, performed on different test samples containing
the same amount of positives and negatives instances (respectively 2000, 4000, 6000, 8000 and 10000 sequences).}
\label{table}
\end{center}

\end{figure}

To compare the quality of the ELF3, HNF1 and MAFK logos we have
computed the Area Under the ROC Curve (AUC) of each model
on a series ground-truth test samples.
For each transcription factor, the test samples consisted 
of 1000, 2000, 3000 4000, and 5000 positive instances, randomly selected 
from the transcription factor dataset, and an equal number of 
negative instances from a control dataset 
(we have used sequences from the ZIC1 dataset).
Since motifs produced by different algorithms have different lengths, 
we have reduced their size down to the size of the shortest one   
by selecting the same 6-8 positions in the logo.
We have defined the score function
to be the maximum of the PWM's log-likelihood 
over all possible positions in the sequence and 
in its reverse complement.
For each model, we have computed the AUC values associated to this 
score function on the five different test samples.
In Figure \ref{table} we report the average of the obtained AUC 
values and the corresponding standard variations.
A possible problem of our AUC-test is the uncertainty about the 
ground-truth test sample, built  on the (probably false) assumption
that all sequences in a dataset come from effectively bounded 
DNA fragments.

\begin{figure}[t]
\centering
\begin{subfigure}{.62\textwidth}
\includegraphics[scale=0.55,trim=30mm 80mm 0 50mm,clip=true]{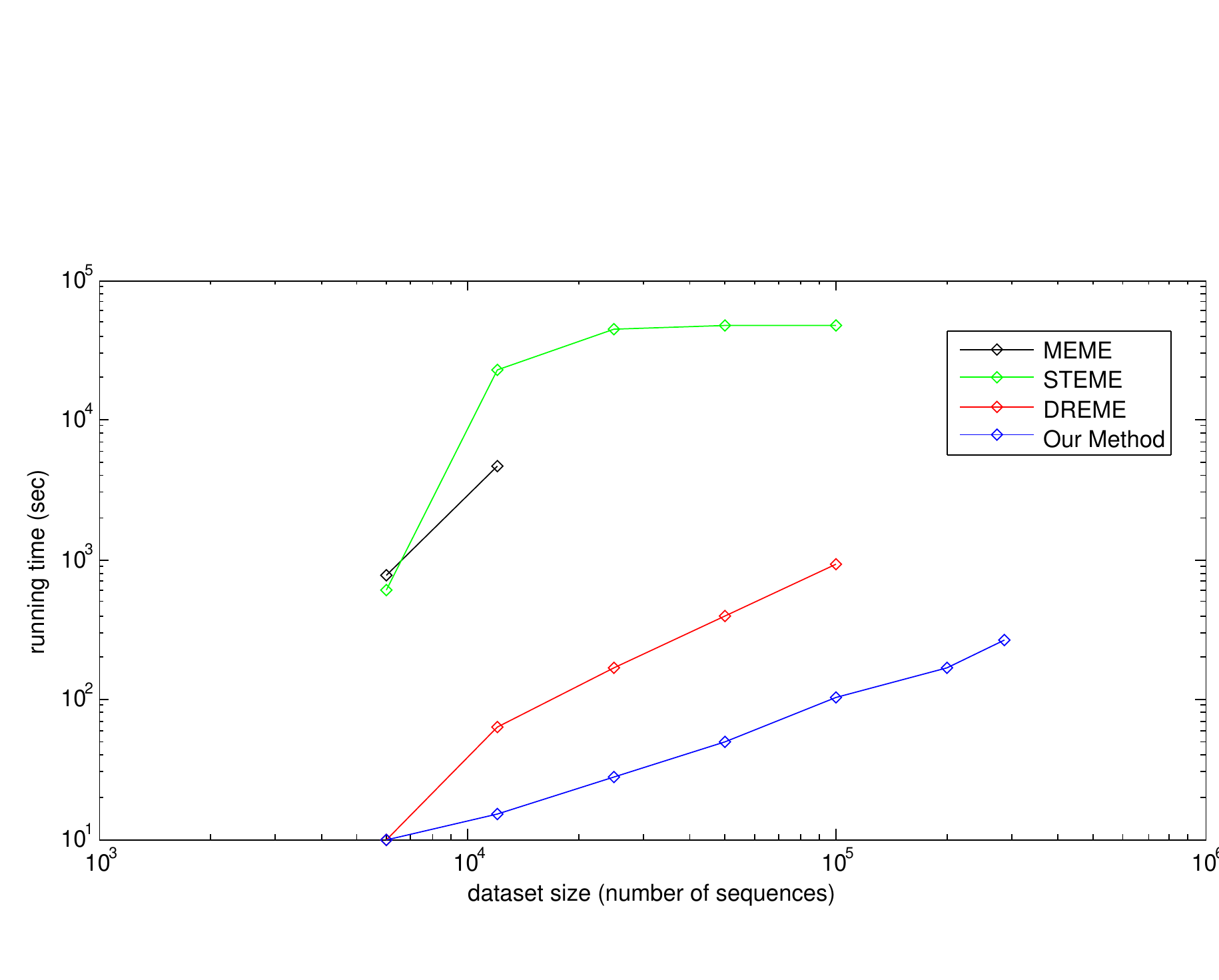}
\end{subfigure}
\begin{subfigure}{.35\textwidth}
\includegraphics[scale=0.5,trim=100mm 70mm 0mm 50mm,clip=true]{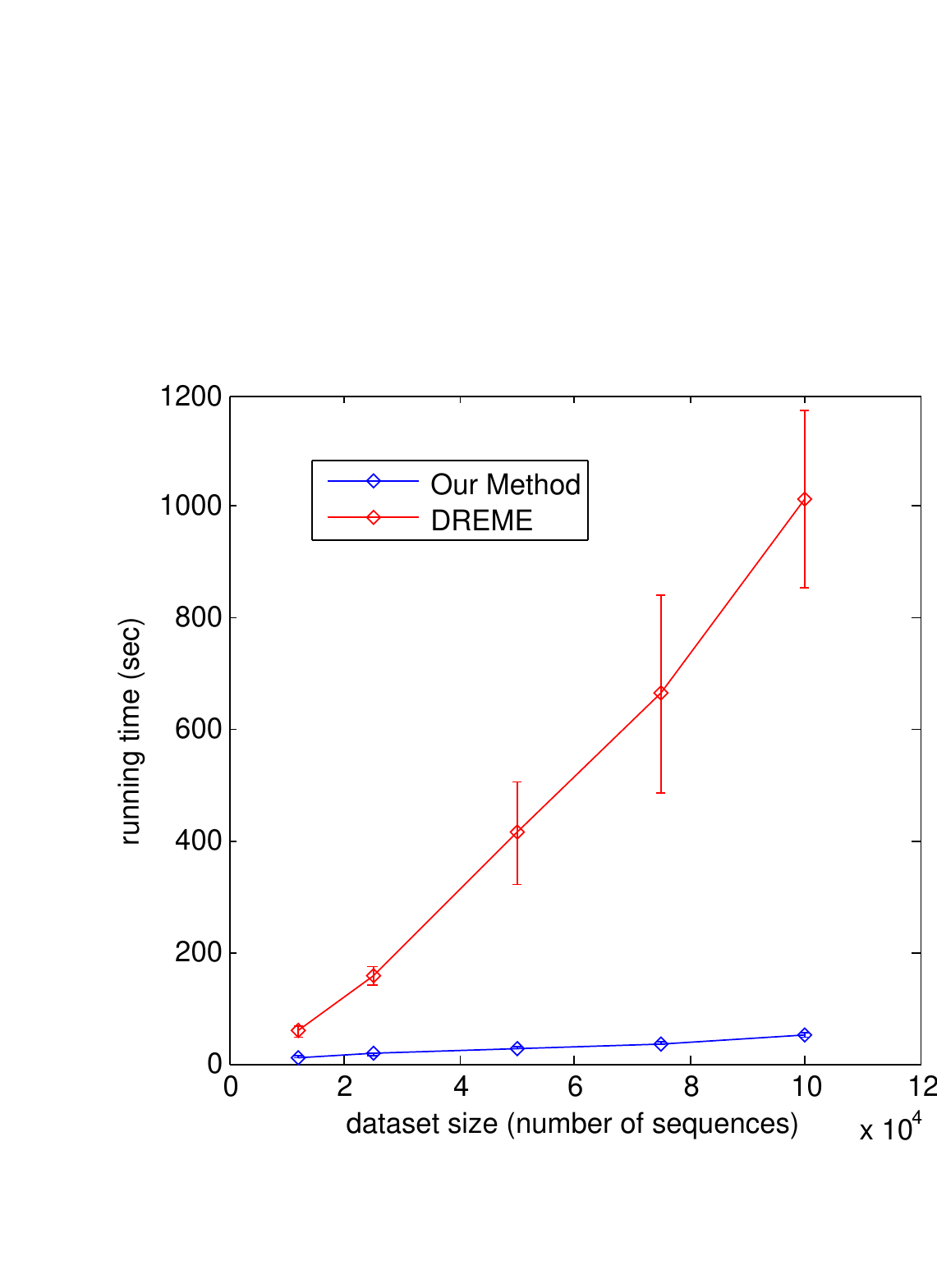}
\end{subfigure}
\caption{Left: Running times for various algorithms 
plotted against the dataset size (all algorithms
ran on the HNF1A dataset of \cite{jolma_dna-binding_2013});
Right: Running times for our algorithm and DREME are plotted against the
size of the datasets. Each point and corresponding errorbar 
represent respectively the average time and the standard deviation 
over three different runs, on the ELF3, HNF1A and 
MAFK datasets of \cite{jolma_dna-binding_2013} respectively.}
\label{time}
\end{figure}

Finally, we have compared the running times of the various algorithms
on datasets of different sizes.
In Figure \ref{time} (left) we show the execution times of DREME, MEME, STEME 
and our algorithm 
on sample containing respectively 3000, 6000, 12500, 25000 and 50000 
randomly selected sequences of the HNF1A dataset.
For all transcription factors and all dataset sizes
our method has been the fastest and DREME the second fastest algorithm.
The running time differences between our method and the others dramatically increase as the size of the dataset grows, see Figure \ref{time}.
We remark the unusual behaviour of STEME whose running times increase very
rapidly for small dataset sizes and reach a plateau at a sample size of 12500.   
MEME has only two values in the plot of Figure \ref{time} because the
algorithm can only support input files containing up to 60000 characters, \emph{i.e.} 3000 sequences
of length 20 bp.
We also report a more careful comparison between the running times of 
DREME and our algorithm 
on datasets coming from three different experiments (ELF3, HNF1A and MAFK).
For five given sample sizes, respectively 6000, 12500, 25000, 
37500 and 50000 sequences, 
we have plotted the average time over
the three different datasets and the corresponding 
standard deviation as errorbars, see \ref{time} (right).   
Unfortunately, we could not find any information about the running times 
of the algorithm used in \cite{jolma_dna-binding_2013}.

\section{Conclusions}
Under the (reasonable) approximation that TF-DNA binding affinities 
are position-independent, the problem of finding the over-represented 
motifs in a set of genome sequences is equivalent to learning a mixture 
of product distributions. 
The inference of mixtures of product distributions is a well known 
problem in computer science, and powerful techniques have been developed 
to solve the problem using spectral decomposition techniques. 
We have applied this idea to the problem of transcription factor binding 
motif discovery and developed an efficient fast motif discovery 
algorithm that computes globally optimal solutions and can support input 
datasets in the order of hundreds of thousand sequences. 
We have tested our method on HT-Selex experimental data and our 
algorithm produces PWM's that match the profiles obtained by the 
state-of-the-art motif finding algorithms, but orders of magnitude faster. 
Future developments include the extension of our method to transcription 
factors binding models that that go beyond the PWM approximation, and 
corresponding theoretical analysis.

\subsubsection*{Acknowledgments}
We would like to thank Anke Wienecke-Baldacchino, 
Merja Heinaniemi, Matthieu Sainlez, Luis Salamanca
and especially C\'edric Laczny
for useful discussions, comments and questions.
This work was supported by  FNR-Luxembourg CORE grant 12/BM/3971381 HIBIO.

\bibliography{my_library.bib}

\appendix
\section{Technical Details}
\label{spectral appendix}

\paragraph{Number of Mixture Components and Grouped Variables}
When Transcription Factor binding affinities are approximated by 
product distributions, motif PWM can
be identified with the conditional probabilities
of a mixture of product distributions and learned using spectral techniques.
We associate the primary motifs in the dataset with the mixture components
of the smallest relative entropy, where the relative entropy is 
computed  with respect to a background distribution
estimated from a control sample.
In the case of noisy data,
besides such primary motifs 
a number of non-specific binding sub-sequences as 
regions containing repeated letters
or low-complexity patterns, can also be over-represented in the dataset.
Moreover, if the size of the sliding window used to compute the 
empirical distributions does not match exactly the size of the target motif, 
strong signals for many shifted versions of the same sub-sequence
should be expected.
We take these effects into account
by defining a preliminary model that includes
a number of extra components,
to be associated with all secondary motifs in the dataset and 
shifted replicates in the sliding window records.
In other words, we learn a mixture with a high and arbitrary
number of components, say $p>15$, and select the 
non-trivial models at the end 
according to their relative entropy computed respect to the background 
distribution.

A limitation of spectral methods is the fact that, by construction, the 
size of the sequence alphabet $d=|{\cal A}|$
bounds from above the number of components
in the mixture, being $d$ the maximum 
number of distinct eigenvalues
that can be obtained from the spectral decomposition of a $d\times d$ matrix.
To overcome this limitation we increase the dimensionality of the alphabet 
by grouping variables.
We can then perform the 
spectral decomposition in a higher dimensional space.
More explicitly,
grouped variables of dimension $d^n$ are obtained by reading 
$n$ contiguous letters as a single character, belonging to the bigger alphabet ${\cal A}^{\otimes n}$.
Given the length  of the sliding window $\ell=3 n $, we define grouped variables of dimension $d^n$ as follows
\begin{equation}
\label{grouping}
 \underbrace{x_1\quad \cdots \quad  x_n}_{x} \qquad \underbrace{x_{n+1} \quad \cdots \quad x_{2 n }}_{y} \qquad  \underbrace{x_{2n +1} \quad \dots\quad  x_{\ell} }_{ z}   
\end{equation}
where $x_i\in {\cal A}$ and $x,y,z\in {\cal A}^{\otimes n}$. 

\paragraph{Sliding Window and Joint Empirical Probabilities}
In the space of grouped variables, given a dataset of sequences ${\cal S}$, the empirical joint distributions are defined by 
\begin{equation}
\label{joint distribution}
[\hat P(x,y,z)]_{i,j,k}~=~\frac{1}{\cal N}\sum_{s\in {\cal S}} \ \sum_{i=1}^{|s|-\ell}  \delta_{s(i:i+n-1),i} \  \delta_{s(i+n:i+2n-1),j}   \ \delta_{s(i+2n:i+\ell-1),k}   \qquad i,j,k=1,\dots,d^n
\end{equation} 
where $[P(x)]_i$ is the probability of observing $x=i$,
${\cal N}$ is a normalization factor,
$\ell$ is the length of the sliding window,
$\delta_{ij}$ is the Kronecker delta function,
and $s(a:b)$, with $1\le a\le b\le |s|$, denotes a sub-sequence of  $s\in {\cal S}$ starting at position $a$ and ending at position $b$ inclusive.

We marginalise \eqref{joint distribution} to obtain the pairwise probability matrices 
\begin{equation}
\label{contraction}
[\hat P(x,y)]_{ij}=
\sum_{a=1}^{d^n} [\hat P(x,y,z)]_{ija} {\bf 1}_{a} , \qquad  [\hat P(x,z)]_{ij}=
\sum_{a=1}^{d^n} [\hat P(x,y,z)]_{iaj} {\bf 1}_{a} , \qquad  i,j=1,\dots,d^n
\end{equation}
where ${\bf 1}=[1,\dots,1]$
and reduce the triple probability tensor to a matrix via the further contraction 
\begin{equation}
\label{P theta}
[\hat P_{\theta}(x,y,z)]_{ij}=  \sum_a [\hat P(x,y,z)]_{iaj} \theta_a  
\end{equation} 
where $\theta$ is an arbitrary vector of dimension $d^n$.

\paragraph{Tensor Spectral Decomposition}

Assuming a mixture of product distributions with $p$ mixture components
over a space of dimension $d^n$,
pairwise probability matrices and triple probability tensors read 
\begin{equation}
\label{mixture}
 P(x,y)~=~ X {\rm diag}(h) Y^T \qquad  P(x,z)~=~ X {\rm diag}(h) Z^T  \qquad [P(x,y,z)]_{ijk}~=~ \sum_{r=1}^p h_r [X]_{ir} [Y]_{jr} [Z]_{kr} 
\end{equation}
where $h=[h_1,\dots,h_p]$ is a vector of mixing weights such that $w\cdot{\bf 1}=1$ and $h_r>0$, for all $r=1,\dots,p$, 
and $X,Y,Z$ are $ d^n \times p$ matrices whose columns sum to one.
As in \eqref{P theta}, we reduce the tensor $P(x,y,z)$ to a matrix 
by contracting its second index to an arbitrary $d$-dimensional vector $\theta$ and 
obtain
\begin{equation}
P_\theta(x,y,z)~=~ X {\rm diag}(h) {\rm diag}(\theta^T Y) Z^T
\end{equation}
where ${\rm diag}(h)$ and ${\rm diag}(\theta^T Y)$ are diagonal matrices whose
$p$ entries are the components of the vectors $h$ and  
$\theta^T Y$ respectively. 
When $d^n>p$ it is convenient to reduce the size of the joint probability matrices
by means of a low-rank approximation as follows.
One assumes the matrices $X,Y$ to have rank $p$ and $h_r>0$ for all $r=1,\dots,p$ and 
defines
\begin{equation}
\label{chang matrix 1}
S_y(\zeta)~=~ V_x^T P_\zeta(x,y,z)  V_z \  \left( V_x^T P(x,z) V_z \right)^{-1}~=~  V_x^T X {\rm diag}(\zeta^T U_y^T Y ) (V_x^T X)^{-1}
\end{equation}
where 
\begin{equation}
[P_\zeta(x,y,z)]_{ij} =  \sum_a [ P(x,y,z)]_{iaj}  \ [\zeta^T U_y^T]_a  
\end{equation}
$\zeta$ is any $p$-dimensional vector and  $U_x,U_y,V_x,V_z$ are the $d^n\times p$ orthogonal matrices 
defined by
\begin{equation}
\hat P(x,y)|_{rank-p}~=~ U_x \ \Sigma_{xy}  \ U_y^T  \ , \qquad  \hat P(x,z)|_{rank-p}~=~V_x \ \Sigma_{xz}  \ V_z^T 
\end{equation}
with $\Sigma_{xy}, \Sigma_{xz}$ being the diagonal matrices containing the $p$ singular values of 
$\hat P(x,y)$, $ \hat P(x,z)$.   
The matrix $S_y(\zeta)$ is called "observable" because it can be 
estimated directly from the sample 
using the empirical matrices $\hat P(x,y,z)$ and $\hat P(x,z)$.
Focusing on the eigenvalues of $S_y(\theta)$, 
relation \eqref{chang matrix 1} implies that it is possible,
under the model assumption  $rank(X) = p = rank(Y)$ and $h_r>0$ 
for all $r=1,\dots p$, to
recover the conditional probabilities $Y$
of a mixture of $p$ hidden components over a space of dimension $d^n$,
via the SVD decomposition of two $d^n\times d^n$ matrices 
and the spectral decomposition
of few $p\times p$ matrices (actually one needs exactly $p$ matrices to obtain $Y$, see below).

\paragraph{Approximate Eigenvalues}
The empirical estimation of \eqref{chang matrix} for a given $p$-dimensional vector $\zeta$ is defined as
\begin{equation}
\label{empirical S}
\hat S_y(\zeta_i)=V_x^T \hat P_\zeta(x,y,z) V_z \  \left( V_x^T \hat P(x,z) V_z \right)^{-1}  , \qquad [\hat P_\zeta(x,y,z)]_{ij} = \sum_a [ \hat P(x,y,z)]_{iaj}  \ [\zeta^T U_y^T]_a  
\end{equation}
When the  empirical distributions are not drawn exactly 
from a mixture of product distributions or the sample size is finite,
the eigenvalues and eigenvectors of \eqref{empirical S}
can contain negative or imaginary parts.
A theorem, \emph{Lemma 4 (Eigenvalue Separation)} in \cite{mossel_learning_2006}, states that, for any $0<\alpha<1$, 
an eigenvalues separation of size $\sim \alpha$ is guaranteed with probability at least $1-\alpha$ 
if the vector $\zeta$ is chosen to be a Gaussian random vector with zero mean and
variance one. 
However, from a more practical point of view, the fact that the
eigenvalues spacing and the failure probability are roughly of the same
order is often source of instabilities \cite{hsu_spectral_2012}.

To increase the stability of our algorithm we choose a set of
distinct random vectors $\zeta_1, \dots \zeta_p$
and perform an approximate joint diagonalisation
of the corresponding nearly-commuting matrices $\hat S_y(\zeta_i)$ \cite{anandkumar_tensor_2012,anandkumar_method_2012}..
Following \cite{corless_reordered_1997},  we compute
the Schur decomposition of a linear combination of the nearly commuting matrices $\hat S_y(\zeta_i)$, \emph{i.e.}
\begin{equation}
\label{schur decomposition}
\hat S_y=\frac{1}{p}\sum_i^{p}\hat S_y(\zeta_i) = Q^T \ \sigma \ Q ,  
\end{equation}
where $Q$ is an orthogonal matrix and $\sigma $ is upper triangular.
The orthogonal matrix $Q$ is then used to compute the 
conditional probabilities matrix $Y$  by choosing $\zeta_i={\bf e}_i$ 
for all $i=1,\dots,p$, 
where ${\bf e}_i$ is a vector with a one in the $i$th coordinate 
and zero otherwise.
For each $i,j=1,\dots p$ we compute 
\begin{equation}
\Lambda^y_{ij}= \left[ Q^T\hat S_y({\bf e}_i) Q\right]_{jj}  
\end{equation}
where $\hat S({\bf e}_i)$ is defined in \eqref{empirical S} 
(with $\zeta_i={\bf e}_i$), 
and then obtain $Y=U_y \Lambda^y$.
 Analogously, we recover $X,Z$ from $X=U_x \Lambda^x$ and $Z=V_z \Lambda^z$ where, 
 for each $i,j=1,\dots p$ we calculate  
\begin{equation}
\Lambda^x_{ij}= \left[ E_x \hat S_x({\bf e}_i) E_x\right]_{jj} \  ,  \qquad 
\Lambda^z_{ij}= \left[ E_z \hat S_z({\bf e}_i) E_y\right]_{jj}  
\end{equation}
with
\begin{eqnarray}
 E_x= W_y^T Y \ , & 
\quad \hat S_x(\zeta_i)=W_y^T \hat P_{x\zeta}(x,y,z) W_z \  \left( W_y^T \hat P(y,z) W_z \right)^{-1} \ , &\quad [\hat P_{x\zeta}(x,y,z)]_{ij} ~=~ \sum_a [ \hat P(x,y,z)]_{aij}  \ [\zeta^T U_x^T]_a\quad  \\
 E_z= V_x^T X  \ ,  &
\quad \hat S_z(\zeta_i) =U_x^T \hat P_{z\zeta}(x,y,z) U_y \  \left( U_x^T \hat P(x,y) U_y \right)^{-1} \ ,&\quad [\hat P_{z\zeta}(x,y,z)]_{ij} ~=~ \sum_a [ \hat P(x,y,z)]_{ija}  \ [\zeta^T V_z^T]_a\qquad 
\end{eqnarray}  
and $W_y,W_z$ defined by
\begin{equation}
\hat P(y,z)|_{rank-p} = W_y \Sigma_{yz} W_z^T 
\end{equation}
where $\Sigma_{yz}$ is the diagonal matrix of singular values of $\hat P(y,z)$. 

It is possible to relate the error on the obtained eigenvalues to the non-commutativity of the nearly commuting empirical matrices $\hat S_y(\zeta_i)$ as follows. 
Letting  $\varepsilon = \max_{i,j} \| \hat S_y(\zeta_i) \hat S_y(\zeta_j)-\hat S_y(\zeta_j) \hat S_y(\zeta_i) \|_F$, 
where $\|A \|_F=\sqrt{{\rm Tr}[A^T A]}$,  and $\tilde T_i=Q^T \hat S_y(\zeta_i) Q $ with $Q$ defined in \eqref{schur decomposition}, it easy to show that, for all $i=1,\dots,p$
\begin{equation}
\label{bound commutator}
\sigma \tilde T_i - \tilde T_i\sigma  = Q^T E_{i} Q  \ , \qquad E_i = \frac{1}{p}\sum_{j=1}^p \left(\hat S_y(\zeta_j) \hat S_y(\zeta_i)-\hat S_y(\zeta_i) \hat S_y(\zeta_j) \right)
\end{equation}
and $\| Q^T E_{i} Q\|_F \le \varepsilon$. 
This implies that $[\tilde T_i]_{j,k}=O(\varepsilon)$ for $j>k$, provided that the separation between the eigenvalues of $\hat S_y$ is $O(1)$ (see \cite{corless_reordered_1997} for more details). 
The error on the eigenvalues is then estimated via the Bauer-Fike theorem \cite{bauer_norms_1960}, using $\varepsilon$ as an upper bound for the norm of the perturbation matrix. More precisely, for all $j=1,\dots,p$ one can find a $k$ such that
\begin{equation}
| \Lambda^y_{ij} - [\zeta_i^T U_y^T Y]_k  | ~\le~ \kappa(U_x^T X)  \  \varepsilon \qquad \forall \ i 
\end{equation}   
where  $\kappa(V)=\|V\| \| V^{-1} \|$.

\paragraph{Higher Dimensional Models and Components Selection}

The output of the spectral decomposition consists 
of three $d^n\times p $ matrices $X,Y,Z$, that contain the 
conditional probability distributions over the $d^n$-dimensional space, respectively for the variables $(x,y,z)$.
For each $r=1,\dots,p$, the columns of $X,Y $ and $Z$
combine to form the higher dimensional PWM's defined by
\begin{equation}
H_r \in [0,1]^{d^n\times 3} \qquad s.t. \left\{ \begin{array}{l} [H_r]_{i1}=X_{ir} , \qquad [H_r]_{i2}=Y_{ir} , \qquad [H_r]_{i3}=Z_{ir}\\ \sum_i[H_r]_{ik} =1 \qquad \forall k\end{array} \right.
\end{equation}
For all $r=1,\dots p$ we compute the relative 
entropy of $H_r$ and choose the models corresponding to the $p_{\rm top}$ smallest values.
The relative entropy is given by   
\begin{equation}
\label{relative entropy}
I(r)=\sum_{i=1}^{d^n} \sum_{k=1}^3  [H_r]_{ik}\log\left(\frac{[H_r]_{ik}}{B_{ik}}\right)
\end{equation} 
for $r=1,\dots,p$, where $B$ is a background distribution defined by  
\begin{equation}
\label{background}
B_{i1}=\sum_{j,k=1}^{d^n}[\hat P_b(x,y,z)]_{ijk} {\bf 1}_j{\bf 1}_k \ , \quad B_{i2}=\sum_{j,k=1}^{d^n}[\hat P_b(x,y,z)]_{jik} {\bf 1}_j{\bf 1}_k\ , \quad B_{i3}=\sum_{j,k=1}^{d^n}[\hat P_b(x,y,z)]_{jki} {\bf 1}_j{\bf 1}_k 
\end{equation}
with ${\bf 1}$ being a $d^n$-dimensional vectors of ones and $\hat P_b(x,y,z)$ 
the empirical triple probability tensor computed using \eqref{joint distribution} for the control dataset. 
In particular, we define the control sample $C$
to be a random selection of DNA fragments bounded 
by a different transcription factor.
This is a good background choice because allows one to exclude a number
of non-specific bounding motifs, expected to be similar in all experiments.

\paragraph{Low Dimensional PWM's and Threshold Selection}

According to their relative entropies, 
$p_{\rm top}$ higher-dimensional models $H^{\rm top}_r$ 
are selected and used to obtain the corresponding  
low-dimensional frequency matrices $h^{\rm top}_r$, 
that contain the binding probability over the original alphabet ${\cal A}$.
Instead of marginalizing the grouped conditional probabilities in each $H^{\rm top}_r$,
we use the high-dimensional models to  
obtain various sub-sequences alignments from the sample ${\cal S}$ and 
then compute the corresponding $h^{\rm top}_r$ by 
counting the occurrences of the $d$-dimensional  
characters at every position $x_1,\dots, x_\ell$.

More precisely, for each model $r=1,\dots, p_{\rm top}$, we assign a 
score to all length-$\ell$ sub-sequences in the dataset ${\cal S}$ 
using the scoring function  
\begin{equation}
f_{\rm alignment}(s_{\ell},r)~=~\log([H^{\rm top}_r]_{s_{\ell}(1:n-1)1})+\log([H^{\rm top}_r]_{s_{\ell}(n:2n-1)2})+\log([H^{\rm top}_r]_{s_{\ell}(2n:\ell)3})
\end{equation}
where $s_{\ell}$ is a sub-sequence of length $\ell$ in ${\cal S}$
and $s_\ell(a:b)$ for $1\le a \le b \le \ell$ is the string 
$[s_\ell(a), s_\ell(a_+1), \dots, s_\ell(b-1),s_\ell(b)]$, 
with $s(i)$ denoting the $i$th element 
of $s_\ell$. 
For $r=1,\dots, p_{\rm top}$, we include in the  $r$th alignment 
the sub-sequences that satisfy $f_{\rm alignment}(s_\ell,r)>\xi$, 
where $\xi$ is a threshold chosen out of a finite set $\{\xi_1,\dots \xi_N\}$
in order to maximise the information content of the model.
More explicitly, for every threshold value $\xi$ in a finite set 
$\{\xi_1,\xi_2,\dots \xi_N\}$, where $N$ is an input parameter,
we compute a $d$-dimensional frequency matrices defined by
\begin{equation}
\label{alignment definition}
[h^{\rm top}_r(\xi)]_{i,j}=\sum_{s_\ell \in S(\xi)} \delta_{i,s(i+j-1)} , \qquad  S(\xi)= \{s_{\ell} \in {\cal S} | f_{\rm alignment}(s_\ell,r) >\xi \}
\end{equation}
and evaluate the information content of the model using
\begin{equation}
\label{information content}
R_r(\xi)=\sum_{k=1}^{\ell} \left[ 2 - (E_r(\xi,k) + \varepsilon_r(\xi,k))\right]   
\end{equation} 
where the entropy $E_r(\xi,k)$ and the small sample correction $\varepsilon_r(\xi,k)$ are
\begin{equation}
E_r(\xi,k)=- \sum_{i=1}^{d} \frac{[h^{\rm top}_r(\xi)]_{i,k}}{\sum_i [h^{\rm top}_r(\xi)]_{i,k}} \log \left(\frac{[h^{\rm top}_r(\xi)]_{i,k}}{\sum_i [h^{\rm top}_r(\xi)]_{i,k}} \right) , \qquad \varepsilon_r(\xi,k)= \frac{d-1}{2 \log(2)} \frac{1}{\sum_i [h^{\rm top}_r(\xi)]_{i,k}} 
\end{equation}
Finally, we define the most informative model to be $h^{\rm top}_{r^*}(\xi^*)$, with
\begin{equation}
 (r^*,\xi^*)= {\rm arg}\ \max_{r,\xi} \ R_r(\xi) , \qquad r=1,\dots,p_{\rm top} , \quad  \xi \in \{\xi_1,\xi_2,\dots \xi_N\}
\end{equation} 
and compute its logo using the application \texttt{weblogo} available at \texttt{http://weblogo.berkeley.edu/} \cite{schneider_sequence_1990}.

Note that in the case of stochastic models, the problem of 
defining an optimal matching threshold given a desired 
minimal statistical significance of the selected instances 
(usually in terms of P-value) 
is an NP-hard problem (see for example \cite{touzet_efficient_2007}).
We herein propose a heuristic solution to this problem.
Improvements to this heuristic go beyond the scope of the current work 
and will be subject of future studies.

\section{Algorithms}
\label{algorithm appendix}
\subsection*{Main algorithm}

{\bf Input:} dataset ${\cal S}$ of sequences over alphabet ${\cal A}$ , 
number of mixture components $p$, 
group size $n$, 
control dataset $C$ ,
threshold values $\xi=\xi_1,\dots, \xi_N$, 
\begin{itemize}
	\item use a sliding window of length $\ell=3  \ n$ to compute $3 n$-mers occurrences from ${\cal S}$ 
	\item convert grouped characters to index and form the sparse tensor $\hat P(x,y,z)$ from a coordinates matrix with rows
	$${\rm index}_x \qquad {\rm index}_ y \qquad  {\rm index}_z \qquad {\rm value}$$
	where $1\le {\rm index}_a \le |{\cal A}|^n$ for $a=x,y,z$
	\item repeat the previous steps using $C$ instead of ${\cal S}$ to obtain
	$\hat P_b(x,y,z)$ and compute
	the background conditional distributions $B$ using \eqref{background}
	\item decompose $\hat P(x,y,z)$ using \emph{Spectral Algorithm} 
	to obtain $H_r$, for $r=1,\dots,p$
	\item for $r=1,\dots p$ compute the relative entropy of $H_r$ using \eqref{relative entropy} 
	\item select the $p_{\rm top}$ components $H^{\rm top}_r$ with smallest relative entropy 
	\item {\bf for} $r=1, \dots,p_{\rm top}$ 
		\begin{itemize}
		\item{\bf for} $\xi=\xi_1,\dots, \xi_N$
			\begin{itemize}
				\item compute frequency matrix $h^{\rm top}_r(\xi)$ using \eqref{alignment definition}
				\item compute the associated information content 
				$R(r,\xi)$ using \eqref{information content} 
			\end{itemize}
			\item[]{\bf end for}
		\end{itemize}
		\item[]{\bf end for}
	\item let $(r^*,\xi^*)={\rm arg} \max_{r , \xi} R(r,\xi)$
	\item compute the logo of $h^{\rm top}_{r^*}(\xi^*)$ using \texttt{weblogo} 	
\end{itemize}

{\bf Output:} Position Weight Matrix of the Transcription Factor binding site and corresponding logo 
 
\subsection*{Spectral Algorithm}
{\bf Input:} coordinates matrix for the $d^n\times d^n \times d^n$ joint probability tensor $\hat P(x,y,z)$, number of mixture components $p$		
\begin{itemize}
\item compute the sparse tensor $\hat P(x,y,z)$ from the coordinates matrix 
and the pairwise probability matrices $\hat P(x,y),\hat P(x,z),\hat P(y,z)$ 
by contracting to ${\bf 1}$ on the third, second and first index 
respectively \footnote{We have used the online available package 
\texttt{MATLAB Tensor Toolbox} \cite{bader_matlab_2012} for all tensors manipulation.}
\item compute the rank$-p$ approximation
\begin{equation}
\hat P(x,y)|_{rank-p}= U_{xy} \Sigma_{xy} V_{xy}^T \ ,\qquad \hat P(x,z)|_{rank-p}=  U_{xz} \Sigma_{xz} V_{xz}^T \ ,\qquad  \hat P(y,z)|_{rank-p}= U_{yz} \Sigma_{yz} V_{yz}^T \
\end{equation} 
\item {\bf for} $i=1,\dots,p$
	\begin{itemize}
	\item  draw a random $p$-dimensional vector $\theta$ and define 
	$v=V_{xy} \theta$
	\item  define $[\hat P_v(x,y,z)]_{ik}~=~ \sum_j [\hat P(x,y,z)]_{ijk}  v_j $
	\item form $M_{v}=U^T_{xz} \hat P_{v}(x,y,z) V_{xz} (U^T_{xz} \hat P(x,z) V_{xz})^{-1} $ 
 	\item let $M=M+M_{v}$
 	\end{itemize}
\item[] {\bf end  for} 	
\item compute the Schur decomposition $M=Q S Q^T$
 \item {\bf for} $r=1,\dots,p$ 
	 \begin{itemize}
 	\item let $v=V_{xy}(:,r)$
 	\item  define $[\hat P_v(x,y,z)]_{ik}~=~ \sum_j [\hat P(x,y,z)]_{ijk}  v_j $
 	\item form $M_{v}=U^T_{xz} \hat P_{v }(x,y,z) V_{xz} (U^T_{xz} \hat P(x,z) 	V_{xz})^{-1}$ 
 	and compute $S_v=Q^T M_{v} Q$
 	\item let $L(r,i)=S_v(i,i)$
 	\end{itemize}
 \item[]{\bf end  for}
 \item let $Y=(V_{xy}L)^{+}$ and normalize the columns of $Y$ to 1
 \item let $E=U_{yz}^TY$
 \item {\bf for} $r=1,\dots,p$ 
	\begin{itemize}
 	\item let $v=U_{xy}(:,r)$ 
 	\item define $ [\hat P_v(x,y,z)]_{jk}~=~ \sum_i [\hat P(x,y,z)]_{ijk} v_i$ 
 	\item form $M_{v}=U^T_{yz} \hat P_{v }(x,y,z) V_{yz} (U^T_{yz} \hat P(y,z) 	V_{yz})^{-1}$ and compute  $D_v=(E)^{-1} M_{v} E$
 	\item let $L(r,i)=D_v(i,i)$
 	\end{itemize}
 	\item[]{\bf end  for}
 	\item let $X=(U_{xy}L)^{+}$ and normalize the columns of $X$ to 1
 	\item let $E=U_{xy}^T X$
 	\item {\bf for} $r=1,\dots,p$ 
	\begin{itemize}
 	\item let $v=V_{xz}(:,r)$ 
 	\item define $[\hat P_v(x,y,z)]_{ij}~=~ \sum_k [\hat P(x,y,z)]_{ijk} v_k $ 
 	\item form $M_{v}=U^T_{xy} \hat P_{v }(x,y,z) V_{xy} (U^T_{xy} \hat P(x,y) 	V_{xy})^{-1}$ and compute  $D_v=(E)^{-1} M_{v} E X$
 	\item let $L(r,i)=D_v(i,i)$
 	\end{itemize}
\item[]{\bf end  for}
 \item let $Z=(V_{xz}L)^{+}$ and normalize the columns of $Z$ to 1

\end{itemize}
{\bf Output:} Position Weight Matrix $H_r\in [0,1]^{d^n\otimes 3}$, for $r=1,\dots p$

\end{document}